\documentclass[times]{elsarticle}
\usepackage[utf8]{inputenc}

\usepackage[margin = 1in]{geometry}
\usepackage{booktabs}
\usepackage{amsmath}
\usepackage{hyperref}
\usepackage{amssymb}
\hypersetup{colorlinks=true, linkcolor=blue, citecolor=blue, urlcolor=blue}
\usepackage[]{natbib}
\usepackage{soul}
\usepackage{multirow}
\usepackage{diagbox}
\usepackage{comment}
\usepackage{float}

\newcommand{\bs}[1]{\boldsymbol{#1}}

\date{\today}
\journal{ }

\begin{document}

\begin{frontmatter}
    \title{Accelerating force calculation for dislocation dynamics simulations}

    \author[label1]{ Rasool Ahmad }
    \ead{rasool@stanford.edu}
    \author[label1]{Wei Cai \corref{cor1}}
    \ead{ caiwei@stanford.edu }
    \address[label1]{Department of Mechanical Engineering,
        Stanford University, CA 94305, USA}
    \cortext[cor1]{corresponding author}

    \begin{abstract}
        Discrete dislocation dynamics (DDD) simulations offer valuable insights into the plastic deformation and work-hardening behavior of metals by explicitly modeling the evolution of dislocation lines under stress. However, the computational cost associated with calculating forces due to the long-range elastic interactions between dislocation segment pairs is one of the main causes that limit the achievable strain levels in DDD simulations. These elastic interaction forces can be obtained either from the integral of the stress field due to one segment over the other segment, or from the derivatives of the elastic interaction energy.  In both cases, the results involve a double-integral over the two interacting segments.  Currently, existing DDD simulations employ the stress-based approach with both integrals evaluated either from analytical expressions or from numerical quadrature.  In this study, we systematically analyze the accuracy and computational cost of the stress-based and energy-based approaches with different ways of evaluating the integrals.  We find that the stress-based approach is more efficient than the energy-based approach.  Furthermore, the stress-based approach becomes most cost-effective when one integral is evaluated from analytic expression and the other integral from numerical quadrature.  For well-separated segment pairs whose center distances are more than three times their lengths, this one-analytic-integral and one-numerical-integral approach is more than three times faster than the fully analytic approach, while the relative error in the forces is less than $10^{-3}$.  Because the vast majority of segment pairs in a typical simulation cell are well-separated, we expect the hybrid analytic/numerical approach to significantly boost the numerical efficiency of DDD simulations of work hardening.
    \end{abstract}

\end{frontmatter}

\section{Introduction}
\label{sec_intro}

Metals and alloys, such as copper, iron, and steel, have always played crucial roles in providing the tools and infrastructures necessary for the continued development of human civilization. The technologically important structural properties of these crystalline materials, including strength, ductility, formability, fracture toughness, creep are directly connected to their plastic deformation behavior under load~\citep{Argon2008, Kubin2013}. Fundamentally, plastic deformation of crystalline materials is governed by motion and evolution of dislocations, linear defects in the crystal lattice, and their interactions with other defects~\citep{Cottrell1965, Anderson2017}. Establishing a quantitative connection between microscopic dislocation evolution and macroscopic mechanical properties has been a long-standing goal of computational materials science.

Atomistic simulations are a widely-used computational tool to probe the fundamental mechanisms in materials behavior~\citep{handbook_materials_modeling}.
Despite their generality and fidelity, the computational cost of atomistic simulations becomes prohibitively high for simulation cell sizes approaching one micron. However, dislocations are known to self-organize into structures with characteristic lengths over several microns. Hence, it is generally expected that a much larger simulation cell (e.g. $>10\,\mu$m) is needed to capture the essential physics of plastic deformation in metals.  These considerations have led to the development of Discrete Dislocation Dynamics (DDD) approach that ignores the atomic-level details and only keeps track of the evolution of dislocations as a line network embedded in an elastic medium~\citep{Giessen1995,Zbib1998,Zbib2002,Arsenlis2007}. Since its origin more than three decades ago, DDD simulation has now emerged as a powerful tool to reveal the fundamental dislocation mechanisms of plastic deformation of crystalline materials.

In DDD simulations, the dislocation lines are discretized into a set of nodes, which are the fundamental degrees of freedom, connected by straight line segments~\citep{Arsenlis2007} or curved splines~\citep{Ghoniem_PRB_2000}. Here, we focus our attention on straight-segment discretization of dislocation network as implemented in ParaDiS~\citep{ParaDiS_site} and other DDD programs ~\citep{Arsenlis2007, Zbib1998,Weygand_MSMSE_2002, Devincre_MSEA_2001}. The force on each node is calculated by considering the long-range elastic interactions between all pairs of dislocation segments and any externally applied load.
The nodal velocities are next computed from the nodal forces using an appropriate dislocation mobility law. The nodes are evolved by numerically integrating the equation of motion. A set of topological operations are then performed to account for atomic-scale dislocation mechanisms such as junction formation, annihilation, and cross-slip. A remeshing step is also applied to maintain good-quality discretization of the dislocation lines as their lengths change. The above steps are repeated until the desired strain level is achieved~\citep{Arsenlis2007,Bulatov2006a,Sills2016}.

One particular feature of DDD simulations is the continuous and steady increase of the degrees of freedom (number of nodes) because of the increase in dislocation density with continued plastic deformation. This increase in degrees of freedom is accompanied by a corresponding increase in the computational cost.  The computational cost associated with force computation is further exacerbated by the fact that for numerical stability the simulation timestep becomes shorter with increasing dislocation density.  This means more computational cycles are needed to reach a given increment of physical time (i.e. a given increment of strain for a constant strain-rate simulation) as the simulation proceeds. The continuous increase in the computational cost with deformation is a major bottleneck that limits the plastic strain accessible by DDD simulations.

Previous attempts to increase the amount plastic deformation accessible by DDD simulations have focused mainly on increasing the timestep taken during one integration step. This is achieved primarily by using an efficient subcycling time integrator~\citep{Sills_MSMSE_2014, Sills_MSMSE_2016}. Furthermore, an implementation of these algorithms on graphical processing units (GPUs) has made it possible to achieve a strain of $\sim 1\%$ in one day wall-clock time for single crystal Cu under a strain rate of $10^{3}$ s$^{-1}$~\citep{Bertin_MSMSE_2019}. Even in these algorithms and implementations, the force calculation remains the most computationally expensive step. Thus, an efficient force calculation algorithm will further enhance the capability of DDD simulations to reach higher strain levels and at lower strain rates.

In DDD simulations, as implemented in ParaDiS, forces arising from the long-range elastic interactions between a pair of segments are described by a non-singular continuum theory of dislocations~\citep{Cai2006}.  The interaction forces on the end nodes of these segments can be obtained using two approaches: (1) from the integral of the Peach-Koehler (PK) force over one segment due to the stress field of the other, and (2) from the derivative of the elastic interaction energy between the two segments with respect to nodal positions.  In the following, we shall refer to the first approach as the stress-based formulation, and the second approach as the energy-based formulation.  The nodal forces from these two formulations do not match each other for a given pair of segments.  But once the contribution from all segment pairs in a dislocation configuration consisting of complete loops are summed together, the total forces on any node obtained from these two formulations agree with each other.

In both the stress-based and energy-based formulations, the nodal force expressions involve a double-integral over the two interacting dislocation segments.  These integrals can either be performed analytically, yielding a close-form expression that can then be evaluated, or be performed numerically (e.g. by Gaussian quadrature).  The current implementation in ParaDiS follows the stress-based formulation in which both integrals have been carried out analytically~\citep{Arsenlis2007}. On the other hand,~\citet{Zbib1998, Zbib2002a, Zbib2002} implement a purely numerical scheme to calculate the forces from stress-based formulation.

In this work, we compare the accuracy and efficiency of various methods to compute the nodal forces using both the stress-based and energy-based formulations.  We confirm that the stress-based formulation leads to more efficient implementations than the energy-based formulation.  Furthermore, we find that the stress-based formulation becomes most efficient when one integral is carried out analytically while the other integral is obtained by numerical quadrature.  For well-separated segment pairs whose center distances are more than three times their lengths, this one-analytic-integral and one-numerical-integral approach is more than three times faster than the fully analytic approach, with the relative error in the forces below $10^{-3}$.  Therefore, we propose to use this hybrid analytic/numerical approach to evaluate the interaction forces for the vast majority of segment pairs beyond a cut-off distance and use the existing fully analytic approach to evaluate forces between segment pairs within the cut-off.  This method should lead to a substantial increase of computational efficiency of DDD simulations with negligible loss of accuracy.

The rest of the paper is organized as follows. Section~\ref{sec:force_calculation} briefly presents the theory behind the interaction forces between two straight dislocation segments. Section~\ref{sec:results} presents the results on the accuracy and computational efficiency of the various numerical implementations. Finally, Section~\ref{sec:conclusion} presents some discussions and conclusive remarks.

\section{Force between two straight dislocation segments}
\label{sec:force_calculation}

In this section, we briefly present the theoretical framework for computing the pair forces between two straight, finite-length, dislocation segments.  In the non-singular continuum elasticity theory of dislocations~\citep{Cai2006}, the Burgers vectors are distributed over a finite region of space instead of concentrated at the dislocation line. A specific isotropic distribution is chosen so that analytic expressions for the stress field of straight dislocation segments and their interaction energies can be obtained, similar to the classical singular elasticity theory of dislocations. The non-singular theory contains an additional core parameter $a$ that characterizes the length-scale of Burgers vector distribution (the singular theory is recovered in the limit of $a\to 0$).
For a finite value of core parameter $a$, the stress field and interaction energies remain finite, and the nodal forces arising from the PK forces due to the stress field are consistent with those from the derivatives of the interaction energy, as long as complete dislocation loops are considered.
As we shall see below, there are multiple approaches to compute the nodal forces from two interacting straight dislocation segments. Although mathematically equivalent, these approaches will lead to implementations with different accuracy and efficiency characteristics.

\begin{figure}[ht!]
    \centering
    \includegraphics[width=\textwidth]{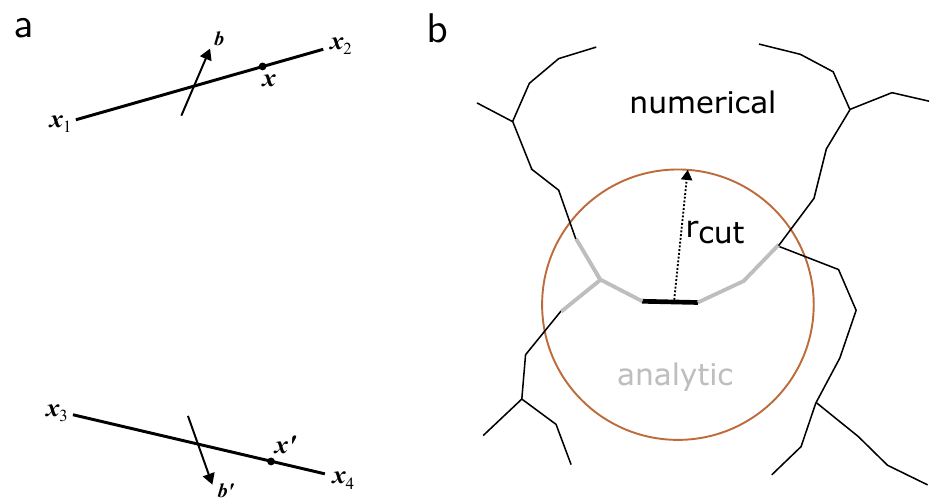}
    \caption{(a) A pair of finite dislocation segments interacting with each other through their elastic fields. The first segment starts at $\bs{x}_1$ and ends at $\bs{x}_2$ with Burgers vector $\bs{b}$, and the second segment starts at $\bs{x}_3$ and ends at $\bs{x}_4$ with Burgers vector $\bs{b}'$. A generic point on segment 1-2 is denoted by $\bs{x}$ and on segment 3-4 by $\bs{x}'$. (b) Schematic illustration of the numeric/analytic hybrid approach to compute force between interacting pairs. Thick solid black dislocation segment in the middle is the one whose interaction is considered with every other dislocation segment. The interactions with dislocation segments (shown in gray) lying within $r_\text{cut}$ (three times the segment length) as well as the interaction with itself is treated with analytic expressions. The interactions with far-away dislocation segments (shown in thin black line) are treated by numerical methods.}
    \label{fig:dislo_pair_network}
\end{figure}

\subsection{Energy-based formulation}
\label{subsec_interaction_energy}

We now present the energy-based formulation of nodal forces. First, let us consider two dislocation loops $C$ and $C'$ with Burgers vector, $\bs{b}$ and $\bs{b}'$, respectively, in an infinite elastic medium with shear modulus $\mu$ and Poisson's ratio $\nu$. The interaction energy between the two loops is given in terms of a double-integral over the two loops as
\begin{equation}
    \begin{aligned}
        E_\text{loop} = & -\frac{\mu}{4\pi} \oint_{C} \oint_{C'} \nabla^2 R_a \, (\bs{b}\times \bs{b}') \cdot (d\bs{x} \times d \bs{x}')+ \frac{\mu}{8\pi} \oint_{C} \oint_{C'} \nabla^2 R_a \, (\bs{b} \cdot d\bs{x})(\bs{b}' \cdot d\bs{x}')            \\
                        & + \frac{\mu}{4\pi(1-\nu)} \oint_{C} \oint_{C'} (\bs{b} \times d\bs{x}) \cdot \bs{T} \cdot (\bs{b}' \times d\bs{x}'),                                                                                                            \\
        \text{where}    &                                                                                                                                                                                                                                 \\
        R_a =           & \sqrt{ \|\bs{x} - \bs{x}'\|^2 + a^2}, \quad \nabla^2 R_a = \frac{2}{R_a} + \frac{a^2}{R_a^3}, \quad \bs{T} =  \frac{\partial^2 R_a}{\partial \bs{x} \partial \bs{x}} = \frac{\partial^2 R_a}{\partial\bs{x}' \partial \bs{x}'}.
    \end{aligned}
    \label{eq:int_eng_loops}
\end{equation}
Here, $\bs{x}$ is point on loop $C$ and $\bs{x}'$ on loop $C'$. We note that there exists an alternate form for the last term in Equation~\ref{eq:int_eng_loops} as derived by \citet{DeWit1959} (in the classical singular continuum theory but easily generalizable to the non-singular theory). The alternate form gives the same result as Equation~\ref{eq:int_eng_loops} as long as the integrals are carried over two closed dislocation loops.  For the rest of this paper, we will continue to use Equation~\ref{eq:int_eng_loops}.

We now consider two straight dislocation segments of finite lengths. As shown in Figure~\ref{fig:dislo_pair_network}(a), the first dislocation segment with Burgers vector $\bs{b}$ starts at $\bs{x}_1$ and ends at $\bs{x}_2$; the second dislocation segment with Burgers vector $\bs{b}'$ starts at $\bs{x}_3$ and ends at $\bs{x}_4$. The interaction energy between the two segments is expressed by simply changing the closed line integrals in Equation~\ref{eq:int_eng_loops} into open line integrals as 
\begin{equation}
    \begin{aligned}
        E_{\rm int} = & -\frac{\mu}{4\pi} \int_{\bs{x}_1}^{\bs{x}_2} \int_{\bs{x}_3}^{\bs{x}_4} \nabla^2 R_a \,(\bs{b}\times \bs{b}') \cdot (d\bs{x} \times d \bs{x}') + \frac{\mu}{8\pi} \int_{\bs{x}_1}^{\bs{x}_2} \int_{\bs{x}_3}^{\bs{x}_4} \nabla^2 R_a \, (\bs{b} \cdot d\bs{x})(\bs{b}' \cdot d\bs{x}') \\
                      & + \frac{\mu}{4\pi(1-\nu)} \int_{\bs{x}_1}^{\bs{x}_2} \int_{\bs{x}_3}^{\bs{x}_4} (\bs{b} \times d\bs{x}) \cdot \bs{T} \cdot (\bs{b}' \times d\bs{x}'),
    \end{aligned}
    \label{eq:int_eng_segments}
\end{equation}
The double-integral in Equation~\eqref{eq:int_eng_segments} can be carried out analytically, yielding closed-form expressions for the interaction energy between two straight dislocation segments~\citep{Cai2006}.  The forces on the four nodes can then be obtained by taking the negative gradient of the interaction energy with respect to the nodal coordinates as
\begin{equation}
    \bs{F}_i = -\frac{\partial E_\text{int}}{\partial \bs{x}_i}, \quad i = 1,2,3,4.
    \label{eq:force_segments}
\end{equation}
Given that the analytic expression of $E_{\rm int}$ is already very complicated, the closed-form expression of $\bs{F}_i$ would be tedious to write down and to implement.  Instead, we can use the automatic differentiation (autograd) tools, widely implemented in modern machine learning packages such as PyTorch, JAX, TensorFlow, etc., to carry out the spatial derivative.  In this work, we use the PyTorch package~\citep{Steiner2019} to perform the automatic differentiation of the analytic energy expression to obtain nodal forces.  We shall call this approach the Energy-Based fully Analytic approach (EB-A), see Table~\ref{tab:method_summary}.

Alternatively, we can perform the double-integral in Equation~\eqref{eq:int_eng_segments} numerically using Gaussian-Legendre quadrature.  The nodal force can then be obtained using autograd.  We shall call this approach the Energy-Based fully Numerical approach (EB-N2), where N2 means both integrals are carried out numerically.  In principle, we can imagine a method in which one of the integrals is carried out analytically and the other one numerically (EB-N1), but we will not examine the performance of this possible implementation in this paper.

\subsection{Stress-based formulation}
\label{sec:stress_formulation}

We now present the stress-based formulation of nodal forces.  In the non-singular elasticity theory, the stress field at a point $\bs{x}$ due to a dislocation loop $C'$ with Burgers vector $\bs{b}'$ is
\begin{align}
    \sigma_{\alpha\beta} (\bs{x}) = \frac{\mu}{8\pi} \oint_{C'}  \frac{\partial^3R_a}{\partial x_i \partial x_p \partial x_p} \left[ b'_m \varepsilon_{im\alpha} dx'_{\beta} + b'_m \varepsilon_{im\beta}dx'_{\alpha}\right] + \frac{\mu}{4\pi(1-\nu)} \oint_{C'} b'_m \varepsilon_{imk} \left(\frac{\partial^3 R_a}{\partial x_i \partial x_\alpha \partial x_\beta} - \delta_{\alpha\beta} \frac{\partial^3 R_a}{\partial x_i \partial x_p \partial x_p} \right)dx'_k
    \label{eq:stress_field_loop}
\end{align}
where $\bs{x}'$ is point on the dislocation loop, $\varepsilon_{ijk}$ is the Levi-Civita symbol, and $\delta_{ij}$ is the Kronecker delta. The stress field due to a finite dislocation segment between $\bs{x}_3$ and $\bs{x}_4$ is obtained simply by converting the closed line integral in Equation~\eqref{eq:stress_field_loop} to an open line integral as
\begin{align}
    \sigma^{3-4}_{\alpha\beta}(\bs{x}) = \frac{\mu}{8\pi} \int_{\bs{x}_3}^{\bs{x}_4}  \frac{\partial^3R_a}{\partial x_i \partial x_p \partial x_p} \left[ b'_m \varepsilon_{im\alpha} dx'_{\beta} + b'_m \varepsilon_{im\beta}dx'_{\alpha}\right] + \frac{\mu}{4\pi(1-\nu)} \int_{\bs{x}_3}^{\bs{x}_4} b'_m \varepsilon_{imk} \left( \frac{\partial^3 R_a}{\partial x_i \partial x_\alpha \partial x_\beta} - \delta_{\alpha\beta} \frac{\partial^3 R_a}{\partial x_i \partial x_p \partial x_p} \right)dx'_k
    \label{eq:stress_field_segment}
\end{align}
The line integral in Equation~\eqref{eq:stress_field_segment} can be carried out analytically, resulting in a closed-form expression for the segment stress~\citep{Cai2006} (see \ref{app:stress_field}).  The nodal forces on the other dislocation segment with endpoints on $\bs{x}_1$ and $\bs{x}_2$ and Burgers vector $\bs{b}$ are then computed by integrating the local PK force due to stress of the segment 3-4 over the segment 1-2 as
\begin{align}
    \bs{F}_1 = \int_{\bs{x}_1}^{\bs{x}_2} \left(\bs{\sigma}^{3-4}(\bs{x}) \cdot \bs{b} \times \bs{t}\right) N_1(\bs{x}) d\bs{x}; \qquad \bs{F}_2 = \int_{\bs{x}_1}^{\bs{x}_2} \left(\bs{\sigma}^{3-4}(\bs{x}) \cdot \bs{b} \times \bs{t}\right) N_2(\bs{x}) d\bs{x},
    \label{eq:nodal_force_segment}
\end{align}
where $\bs{t} = ({\bs{x}_2 - \bs{x}_1})/{\| \bs{x}_2 - \bs{x}_1 \|}$ is the unit tangent vector and  $N_1(\bs{x})$ and $N_2(\bs{x})$ are the linear shape functions of the dislocation segment 1-2. $N_1(\bs{x}_1) = 1$, $N_1(\bs{x}_2) = 0$ and $N_2(\bs{x}_1) = 0$, $N_2(\bs{x}_2) = 1$.  The line integral in Equation~\eqref{eq:nodal_force_segment} can also be integrated analytically, resulting in a closed-form expression for the nodal forces~\citep{Arsenlis2007}.  This is the approach implemented in ParaDiS, and we shall call it the Stress-Based fully Analytic approach (SB-A).

Alternatively, we can use the analytic expression of the segment stress, but evaluate the integral in Equation~\eqref{eq:nodal_force_segment} using Gaussian-Legendre quadrature.  We shall call this hybrid approach SB-N1, where N1 indicates that one of the two line integrals is evaluated numerically.  Furthermore, we can also evaluate both line integrals in Equations~\eqref{eq:stress_field_segment} and \eqref{eq:nodal_force_segment} numerically, and we shall call the approach SB-N2.  The different methods described above are summarized in Table~\ref{tab:method_summary}, and their accuracy and computational efficiency will be compared in the next section.

\begin{table}
    \centering
    \caption{Descriptions of the various methods considered here for evaluating nodal forces due to elastic interaction between dislocation segment pairs. The methods are characterized as being either energy-based (EB) or stress-based (SB), and how the two line-integrals are carried out: both integrals analytic (A), one integral analytic and one integral numeric (N1), or both integrals numeric (N2). }
    \begin{tabular}{l | c | c}
        \toprule
                                  & Energy-based formulation & Stress-based formulation \\
        \midrule
        {Both integrals analytic} & EB-A                     & SB-A                     \\
        \midrule
        One integral analytic     & \multirow{2}{*}{-}       & \multirow{2}{*}{SB-N1}   \\
        One integral numeric      &                          &                          \\
        \midrule
        {Both integrals numeric}  & EB-N2                    & SB-N2                    \\
        \bottomrule
    \end{tabular}
    \label{tab:method_summary}
\end{table}

\section{Results}
\label{sec:results}

In this section, we compare the accuracy and computational efficiency of the various methods for calculating the nodal forces due to elastic interaction between two straight dislocation segments of finite lengths.  All methods are implemented in Python.  The autograd tool in the PyTorch library is used for differentiation of the energy function to obtain nodal forces in energy-based approaches.

To construct the test cases, we randomly generate 8,000 pairs of dislocation segments.  Each segment has a randomly chosen Burgers vectors and random line orientations, but a fixed length of 2 nm.  The separation between midpoints of the two segments is a random number uniformly distributed between 6.0 and 30.0 nm.  The elastic medium has the shear modulus of $\mu = 50$ GPa and Poisson's ratio $\nu = 0.3$. The dislocation core parameter is chosen to be $a = 0.01$ nm.

\subsection{Energy-based methods: EB-A vs EB-N2}
\label{subsec:energy_force}

\begin{figure}[ht!]
    \centering
    \includegraphics[width=\textwidth]{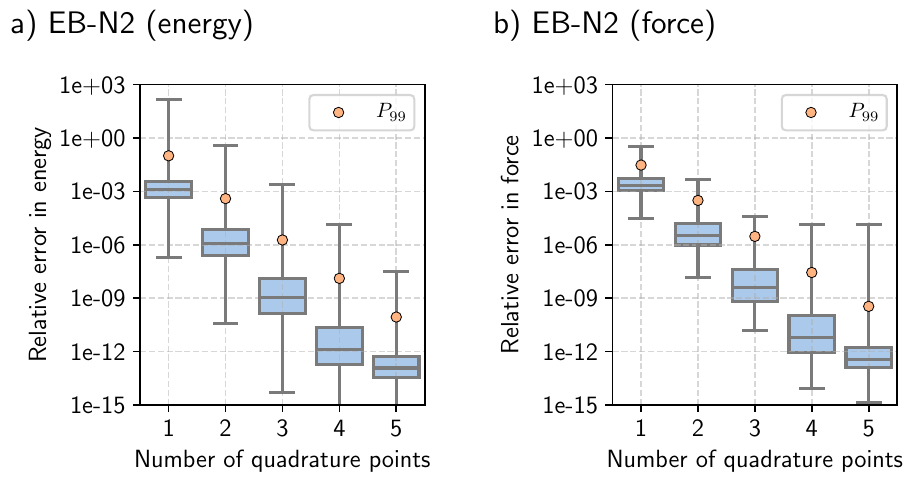}
    \caption{Boxplot of the relative errors of the EB-N2 method in (a) interaction energies and (b) nodal forces between two dislocation segments as a function of the number of quadrature points. The orange dots on each box whisker denotes the $99^{\rm th}$ percentile of the relative error data.  }
    \label{fig:energy_segments}
\end{figure}

Here we compare the interaction energy and forces between two straight segments using the EB-A and EB-N2 methods.  The results from the EB-A method is considered to be exact, and based on which the error of the EB-N2 method is computed.  Figure~\ref{fig:energy_segments}(a) shows that the error in the interaction energy computed by the EB-N2 method decreases exponentially fast with the number of quadrature points on each segment.  The boxplot shows that at each chosen number of quadrature points, there is a significant spread of relative errors among the randomly generated segment pairs.

Figure~\ref{fig:energy_segments}(b) shows the relative error in the forces on the four nodes as a function of the number of quadrature points.  The relative error in forces is computed from the magnitude of the force difference between EB-N2 and EB-A methods divided by the magnitude of the force computed by the EB-A method.  The nodal forces computed by the EB-N2 method also converge to the exact values exponentially fast (in most cases) with an increasing number of quadrature points.  Furthermore, only 3 quadrature points are enough to bring the maximum relative error in force down to below $10^{-4}$, i.e. 0.01\%.  In addition, as shown by the whiskers and the $99^\text{th}$ percentile marks, in the vast majority of cases, the errors are orders of magnitude lower than the maximum error. For instance, using 3 quadrature points, in 75\% of the cases, the relative errors in force lie below $10^{-7}$, and in 99\% of the cases, the relative errors lie below $10^{-5}$.

\begin{table}[ht]
    \centering
    \caption{Time (in seconds) taken by the different methods to evaluate the force due to the elastic interaction between a pair of dislocation segments.}
    \begin{tabular}{c | c c c c c}
        \toprule
        \multirow{2}{*}{Method} & EB-A               & EB-N2                 & SB-A               & SB-N1                 & SB-N2                 \\
                                &                    & (3 quadrature points) &                    & (3 quadrature points) & (3 quadrature points) \\
        \midrule
        Time (s)                & $4.2\times10^{-1}$ & $1.4\times10^{-2}$    & $4.7\times10^{-4}$ & $1.5\times10^{-4}$    & $2.4 \times 10^{-4}$  \\
        \bottomrule
    \end{tabular}
    \label{tab:comp_time}
\end{table}

\begin{figure}
    \centering
    \includegraphics[width=0.5\textwidth]{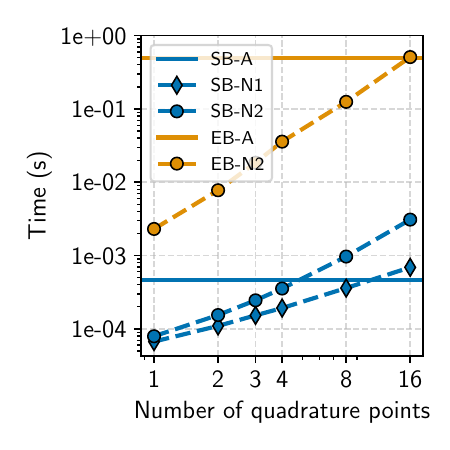}
    \caption{Wall-clock time to determine nodal forces for one segment pair as a function of number of quadrature points for methods SB-N1, SB-N2 and EB-N2. The computational time is computed by averaging over 10000 dislocation segment pairs. Data presented in blue correspond to SB-N1 (diamond marker) and SB-N2 (circular marker) and that in orange correspond to EN2. The time taken by analytic evaluation of energy-based (EB-A) and stress-based (SB-A) forces are also shown in solid horizontal lines for comparison.}
    \label{fig:time}
\end{figure}

We now compare the computation times of EB-A and EB-N2 methods to calculate the interaction forces between a pair of dislocation segments.  The computational time for EB-A and EB-N2 methods using 3 quadrature points are listed in Table~\ref{tab:comp_time}.  These data are also plotted in Figure~\ref{fig:time}, together with the computational time for the EB-N2 methods as a function of the number of quadrature points on each segment.  All these time data are the averages of 10000 different calculations on a CPU machine. The computation time for the EB-N2 method scales quadratically with the number of quadrature points due to the double line integral involved in Equation~\ref{eq:int_eng_segments}.  As shown in Figure~\ref{fig:time}, even when 10 quadrature points are used on each segment, the EB-N2 method is still more efficient than the EB-A method.  On the other hand, Figure~\ref{fig:energy_segments}(b) shows that 3 quadrature points are already sufficient to bring the relative error of the EB-N2 method to below $10^{-4}$.  Therefore, for dislocation pairs where the separation between the midpoints of the segments is more than three times the length of the segments, the EB-N2 method (using 3 quadrature points) is considered to be sufficiently accurate and is 30 times faster than the EB-A method (see Table~\ref{tab:comp_time}).

\subsection{Stress-based methods: SB-A vs SB-N1 vs SB-N2}
\label{subsec:stress_force}

Here we compare the interaction forces between two straight segments using the SB-A, SB-N1 and SB-N2 methods.  The tests are performed on the same set of segment pairs as those used in Section~\ref{subsec:energy_force}.  The results from the SB-A method are considered to be exact, from which the errors of the SB-N1 and SB-N2 methods are computed.

\begin{figure}
    \centering
    \includegraphics[width=\textwidth]{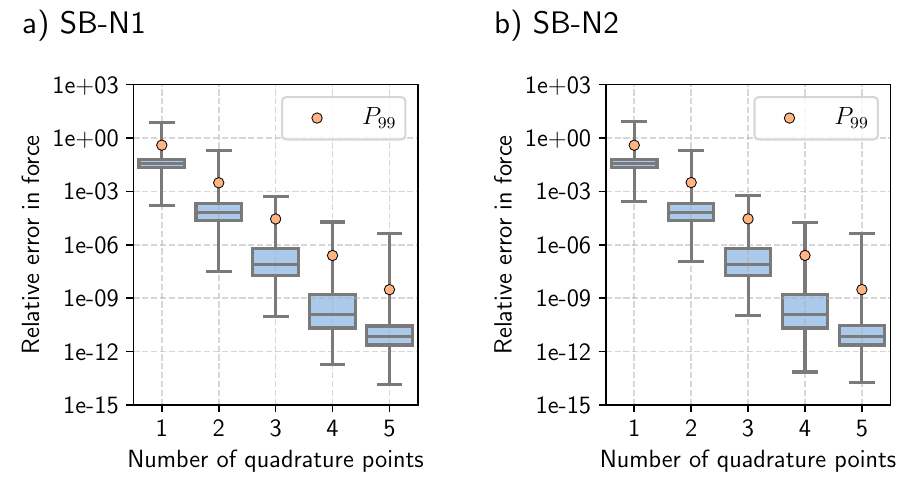}
    \caption{Boxplot of the relative errors in numerically computed nodal forces between two dislocation segments as a function of the number of quadrature points: (a) SB-N1 method, (b) SB-N2 method. Orange dots in each whisker denote $99^{\rm th}$ percentile of relative error data. The relative error in forces is computed by subtracting the analytically computed force from the numerical value and then dividing its magnitude by the corresponding analytic force magnitude.}
    \label{fig:force_stress_segments_cutoff_2}
\end{figure}

Figure~\ref{fig:force_stress_segments_cutoff_2} (a) shows that the results from the SB-N1 method converges exponentially fast with increasing number of quadrature points.  Using 3 quadrature points, the maximum relative error of the SB-N1 method is already less than $10^{-3}$, i.e. $0.1\%$, which is considered sufficiently small. In fact, the vast majority of the relative errors are orders of magnitude lower than the maximum error. For instance, using 3 quadrature points, in 75\% of the cases, the relative errors in force lie below $10^{-6}$, and in 99\% of the cases, the relative errors lie below $10^{-4}$. Figure~\ref{fig:force_stress_segments_cutoff_2} (b) presents the results for SB-N2. We again see that the error decreases exponentially with the number of quadrature points. The error values are almost the same as those of SB-N1, indicating that for far-enough segments, the error is dominated by the numerical integration of PK force, and the contribution of numerical evaluation of stress field does not significantly increase the total error.

The computational times for force evaluation for one pair of segments taken by the SB-A, SB-N1 (3 quadrature points) and SB-N2 (3 quadrature points) methods are given in Table~\ref{tab:comp_time}. Figure~\ref{fig:time} also plots the computational time for the SB-N1 and SB-N2 methods as a function of the number of quadrature points.  The computational time for the SB-N1 method scales linearly with the number of quadrature points, given that only one integral is evaluated numerically. The SB-N1 method is more efficient than the SB-A method for up to 8 quadrature points.  However, Figure~\ref{fig:force_stress_segments_cutoff_2} shows that 3 quadrature points are already sufficient to bring the relative error of the SB-N1 method to below $10^{-3}$.  The computational time of SB-N2 method scales quadratically with the number of quadrature points due to numerical evaluation of both the stress integral, Equation~\eqref{eq:stress_field_segment}, and the integral of PK force, Equation~\eqref{eq:nodal_force_segment}. Even if the computation time of SB-N2 is similar to that of SB-N1 for 1 quadrature point, the quadratic scaling of SB-N2 method makes it less efficient than SB-N1 method for 2 or more quadrature points. Thus, SB-N1 method is more computationally efficient than SB-N2 method.  Therefore, for dislocation pairs where the separation between the midpoints of the segments is more than three times of the length of the segments, the SB-N1 method (using 3 quadrature points) is considered to be sufficiently accurate and is more than three times faster than the SB-A method (see Table~\ref{tab:comp_time}).

Based on Table~\ref{tab:comp_time}, Figure~\ref{fig:energy_segments} and Figure~\ref{fig:force_stress_segments_cutoff_2}, we conclude that the most efficient way to evaluate the forces from a pair of straight dislocation segments is to use the SB-N1 method (with 3 quadrature points) when the two segments are well-separated, i.e. the distance between their midpoints is more than three times the segment lengths.  For segments that are closer than this cut-off distance, the SB-A method is a good choice.  Given that in a DDD simulation, the vast majority of segment pairs are well separated, this combined approach using SB-N1 and SB-A methods based on distances is expected to be much more efficient than using the SB-A method alone.

\subsection{Dislocation loops}

\begin{figure}[ht!]
    \centering
    \includegraphics[width=0.7\textwidth]{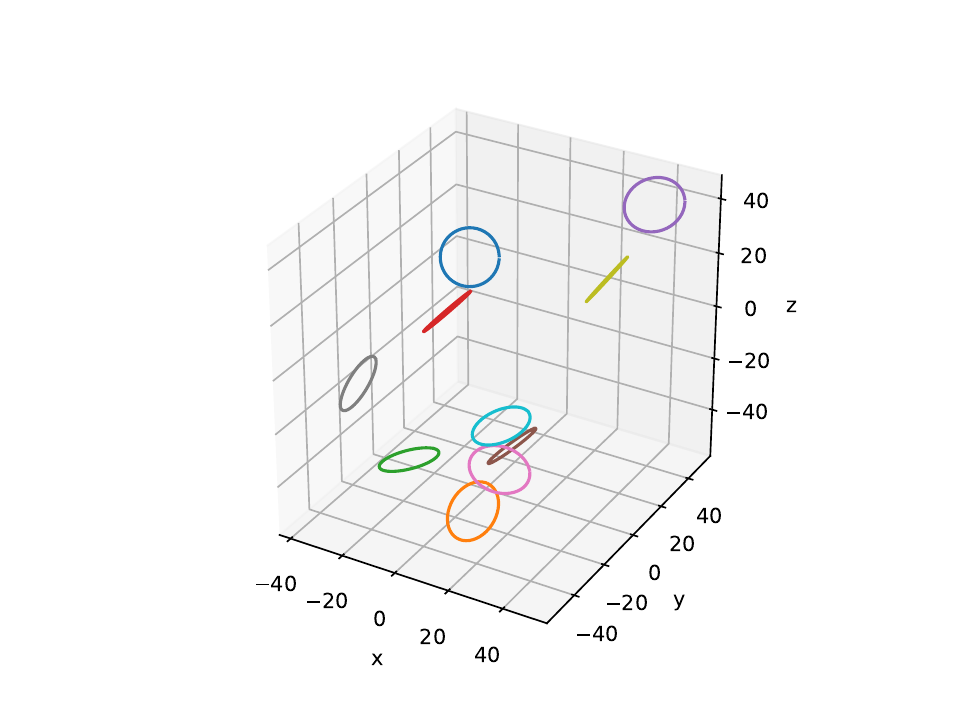}
    \caption{Configuration of ten circular dislocation loops randomly oriented in a three-dimensional infinite linear isotropic elastic medium. Each dislocation loop is discretized into 45 segments, and has Burgers vector of unit magnitude and random orientation.}
    \label{fig:disl_loop}
\end{figure}

We now compare the accuracy and efficiency of different methods in computing the total nodal forces in a scenario that resembles that of a DDD simulation.  To this end, we consider ten circular dislocation loops, each with a radius of 10 nm, randomly oriented in an infinite isotropic linear-elastic medium, as shown in Figure~\ref{fig:disl_loop}.  Each dislocation loop is discretized into 45 segments. The length of each segment is $\approx$ 1.4 nm. The Burgers vectors of the ten loop are chosen to be of unit magnitude and randomly oriented.  Other parameters ($\mu$, $\nu$, $a$) are the same as those in the previous sections.  The force on every node is the result of the interaction between every segment with all segments (including itself).  Since the configuration contains only closed dislocation loops, the forces computed by the energy-based methods are expected to agree with those computed by the stress-based methods.

The forces computed entirely from the SB-A method is considered to be exact, based on which the errors of other methods are computed.  (The maximum relative error of the EB-A method is less than $10^{-3}$.) In order to balance accuracy and efficiency, we consider combined analytic/numerical approaches in which EB-A or SB-A is used for well-separated segment pairs (with center distances greater than three times the segment length), while EB-N2, SB-N1 or SB-N2 is used for the remaining segment pairs. For this test case, this means that essentially the analytic method (EB-A or SB-A) is used only for interactions between a segment with itself and with its four (nearest and next nearest) neighbors.

\begin{figure}[ht!]
    \centering
    \includegraphics[width=0.9\textwidth]{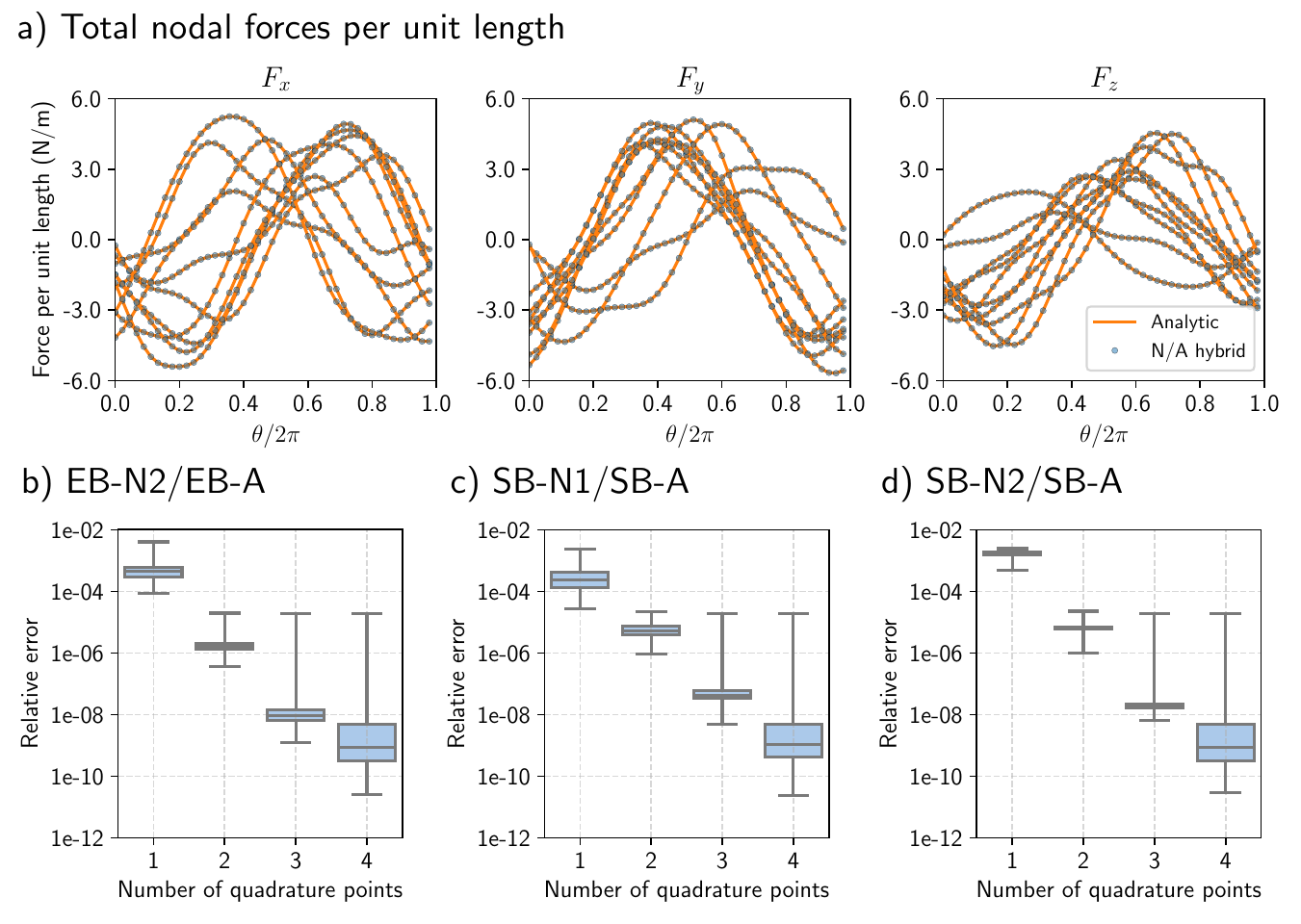}
    \caption{Comparison of analytic and analytic/numerical total nodal force per unit length for the ten dislocation loops shown in Figure~\ref{fig:disl_loop}. (a) Variation of total nodal forces per unit length as a function of angular position of nodes in the dislocation loops. Ten different curves are shown, one for each dislocation loop. Solid orange curves are results from the SB-A method; blue dots are obtained from the analytic/numeric hybrid  where nearby segments (with separation less than three times of the segment length) are handled by the SB-A method and far-away segments by the SB-N1 method (using 3 quadrature points). Both stress and energy formulations lead to the same plot as shown here. Boxplots of distribution of relative errors in forces as a function of quadrature points for the cases of (b) EB-N2/EB-A, (c) SB-N1/SB-A hybrid, and (d) SB-N2/SB-A schemes. Relative error in forces is defined as the ratio of the magnitude of the difference between hybrid and analytic forces to the magnitude of the analytic forces. }
    \label{fig:disl_loop_nodal_forces}
\end{figure}

Figure~\ref{fig:disl_loop_nodal_forces}(a) plots the forces per segment length on all the nodes using the SB-A method and the combined SB-N1/SB-A method (with 3 quadrature points).  The differences between the two methods are too small to be seen in this figure.  Figure~\ref{fig:disl_loop_nodal_forces} (b) shows the relative error in nodal forces using the combined EB-N2/EB-A method as a function of the number of quadrature points.  The maximum relative error is less than $10^{-2}$ even with 1 quadrature point.  For the vast majority of the nodes, the relative error in the nodal forces decays rapidly with the number of quadrature points.

Figure~\ref{fig:disl_loop_nodal_forces} (c) shows the relative error in nodal forces using the combined SB-N1/SB-A method as a function of the number of quadrature points.  The relative error also decreases exponentially with increasing number of quadrature points for the vast majority of the nodes.  The maximum relative error seems to remain stagnant with increasing number of quadrature points when it is 2 or more, at a value below $10^{-4}$, which is considered small enough.  Figure~\ref{fig:disl_loop_nodal_forces} (d) shows the relative error in nodal forces using the combined SB-N2/SB-A method as a function of the number of quadrature points.  The overall behavior is quite similar to that of the SB-N1/SB-A method shown in Figure~\ref{fig:disl_loop_nodal_forces} (c).

The time to evaluate the nodal forces using different methods are given in Table~\ref{tab:comp_time_loop}. The analytic evaluation of forces by SB-A method takes around 50 seconds, while the combined SB-N1/SBA method 3-point quadrature is the fastest and takes 17 seconds.  The combined SB-N1/SB-A method is both highly accurate and about three times as fast as the SB-A method.

\begin{table}[ht]
    \centering
    \caption{Time (in seconds) taken by the different methods to evaluate the nodal forces for the ten dislocation loops.}
    \begin{tabular}{c | c c c c c}
        \toprule
        \multirow{2}{*}{Method} & EB-A             & EB-N2                 & SB-A            & SB-N1                 & SB-N2                 \\
                                &                  & (3 quadrature points) &                 & (3 quadrature points) & (3 quadrature points) \\
        \midrule
        Time (s)                & $4.1\times 10^4$ & $2.1\times10^3$       & $4.9\times10^1$ & $1.7\times10^{1}$     & $2.8 \times 10^{1}$   \\
        \bottomrule
    \end{tabular}
    \label{tab:comp_time_loop}
\end{table}

\section{Discussion and conclusion}
\label{sec:conclusion}

In this work, we compare different ways of computing nodal forces in a DDD simulation using straight dislocation segments.  The methods differ in their theoretical formulation, i.e. either energy-based (EB) or stress-based (SB), as well as how the line-integrals are carried out, i.e. either analytically or by numerical quadrature.  We observe that a combined approach, where interaction forces due to well-separated segments (with center distances more than three times segment length) are computed using SB-N1 and forces due to other segments are computed using SB-A, can be both highly accurate (relative error less than $10^{-3}$) and significantly faster than using SB-A alone.  Therefore, we recommend using such a combined approach for nodal force calculations in DDD simulations.  In a DDD simulation in which the interactions between $N$ segments need to be explicitly accounted for, the number of well-separated segment pairs scales as $\mathcal{O}(N^2)$, while the closely-spaced segment pairs scale as $\mathcal{O}(N)$.  Therefore, the faster SB-N1 method would be used in most cases in place of the exact but slower SB-A method, in the limit of large $N$.  The energy-based methods, unfortunately, are significantly slower than the stress-based methods, most likely due to the need of taking autograd of relatively complicated energy functions.

We note that all benchmark tests in this work are performed using Python codes running on CPUs.  The observed numerical accuracy of the methods (e.g. convergence rate with respect to number of quadrature points) is expected to be generally applicable to implementations using other programming languages and computing platforms.  While we expect the tests here also provide general insights into the relative efficiency of different methods, the exact ratio of computational time between methods can change if they are implemented in a different language (e.g. C language) or running on Graphical Processing Units (GPUs). More work is needed to determine how much speedup can be gained in DDD simulations of work hardening, by applying the methods developed here to C/CPU and Cuda/GPU implementations of ParaDiS.

In conclusion, this work shows that the most computationally intensive part of DDD simulations can be sped up by using more efficient methods for force evaluations between well-separated segment pairs while maintaining high accuracy.  This finding is likely to significantly expand the capability of large-scale DDD simulations of work hardening in metals at reaching higher strains and under lower strain rates.


\appendix

\section{Stress field of a straight dislocation segment}
\label{app:stress_field}

\begin{figure}
    \centering
    \includegraphics[width=0.5\linewidth]{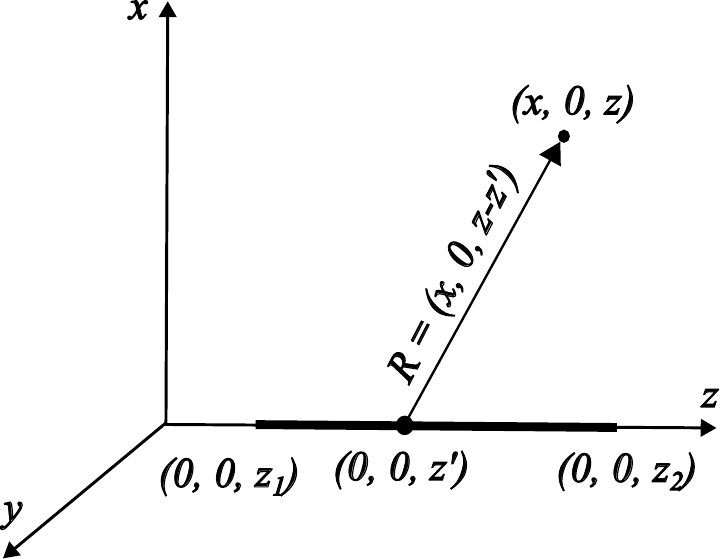}
    \caption{Coordinate system used to describe the stress field on a field point $(x, 0, z)$ due a straight dislocation segment lying along $\bs{z}-$axis from $(0,0,z_1)$ to $(0,0,z_2)$.}
    \label{fig:coord_sys}
\end{figure}

We present the stress field of a finite straight dislocation segment in the framework of the non-singular elasticity theory of dislocations\citep{Cai2006}. The expressions are presented in a coordinate-dependent form. We assume that the dislocation segment with Burgers vector $\bs{b}'$ lies along the $\bs{z}-$axis and extends from $(0,0,z_1)$ to $(0,0,z_2)$ as shown in Figure~\ref{fig:coord_sys}. We determine the stress at the field point $\bs{x} = (x, 0, z)$ which lies entirely in $\bs{x}-\bs{z}$ plane. The vector $\bs{R}=(x, 0, z-z')$ connects a point on the dislocation segment $(0,0,z')$ to the field point $(x,0,z)$.

The stress field, Equation~\eqref{eq:stress_field_segment}, in this special coordinate system can be expressed as
\begin{equation}
    \begin{aligned}
        \frac{\sigma_{xx}}{\sigma_0} & = b'_y \int_{z_1}^{z_2} \left[\frac{3x^3}{R_a^5} - \frac{x}{R_a^3} \left(1-3\left[\frac{a}{R_a}\right]^2\right)\right] dz',                                                                                                                  \\
        \frac{\sigma_{yy}}{\sigma_0} & =  b'_y \int_{z_1}^{z_2}  \frac{x}{R_a^3} \left(1+3\left[\frac{a}{R_a}\right]^2\right) dz',                                                                                                                                                  \\
        \frac{\sigma_{zz}}{\sigma_0} & = b'_y \int_{z_1}^{z_2} \left[\frac{3x(z-z')^2}{R_a^5} - \frac{x}{R_a^3}\right]dz'- b'_y(1-\nu) \int_{z_1}^{z_2} \frac{x}{R_a^3} \left(2 + 3\left[\frac{a}{R_a}\right]^2\right)dz',                                                          \\
        \frac{\sigma_{xy}}{\sigma_0} & = b'_x \int_{z_1}^{z_2} \frac{x}{R_a^3} dz',                                                                                                                                                                                                 \\
        \frac{\sigma_{xz}}{\sigma_0} & = b'_y \int_{z_1}^{z_2} \left[\frac{3(z-z')x^2}{R_a^5} - \frac{z-z'}{R_a^3}\right]dz' + \frac{b'_y(1-\nu)}{2} \int_{z_1}^{z_2} \frac{z-z'}{R_a^3}\left(2 + 3 \left[\frac{a}{R_a}\right]^2\right) dz',                                        \\
        \frac{\sigma_{yz}}{\sigma_0} & = b'_x \int_{z_1}^{z_2} \frac{z-z'}{R_a^3} \left[1 + \frac{1-\nu}{2} \left(2 + 3\left[\frac{a}{R_a}\right]^2\right)\right] dz' - \frac{b'_z(1-\nu)}{2} \int_{z_1}^{z_2} \frac{x}{R_a^3}  \left(2 + 3\left[\frac{a}{R_a}\right]^2\right) dz', \\
    \end{aligned}
    \label{eq:stress_field_diff_ccord}
\end{equation}
where
\begin{equation}
    \sigma_0 = \frac{\mu}{4\pi(1-\nu)}, \qquad R_a = \sqrt{x^2 + (z-z')^2 + a^2}
\end{equation}
The following identities are used to derive the above equations,~\eqref{eq:stress_field_diff_ccord}, from Equation~\eqref{eq:stress_field_segment}
\begin{equation}
    \begin{aligned}
        \frac{\partial^3R_a}{\partial x_i \partial x_j \partial x_k} & = \frac{3x_ix_jx_k}{R_a^5} - \frac{x_i \delta_{jk} + x_j \delta_{ik} + x_k \delta_{ij}}{R_a^3} \\
        \frac{\partial}{\partial x_i} \nabla^2R_a                    & = -\frac{x_i}{R_a^3}\left(2 + 3\left[\frac{a}{R_a}\right]^2\right)
    \end{aligned}
\end{equation}

The integrals in Equation~\eqref{eq:stress_field_diff_ccord} can be evaluated either exactly~\citep{Cai2006} and used in SB-N1 method in the main text, or by numerically using Gauss-Legendre quadrature scheme which is used in SB-N2 method in the main text.

The closed form integral of Equation~\eqref{eq:stress_field_diff_ccord} is expressed as the difference
\begin{equation}
    \sigma_{ij} = \bar{\sigma}_{ij}(z'=z_2) - \bar{\sigma}_{ij}(z'=z_1).
\end{equation}

The stress field $\bar{\sigma}_{ij}(z')$ can be expressed in several equivalent forms, and for numerical stability a particular form should be used depending on the position of the field point relative to the dislocation segment~\citep{Cai2006}. If the field point is located left to the dislocation segment, i.e. $z < z_1 < z_2 $, the following form 1 should be used
\begin{equation}
    \begin{aligned}
        \frac{\bar{\sigma}_{xx}}{\sigma_0} & =  \frac{b'_y x}{R_a(R_a + \lambda)} \left[1 - \frac{x^2 + a^2}{R_a^2} - \frac{x^2 + a^2}{R_a(R_a + \lambda)}\right],                                                               \\
        \frac{\bar{\sigma}_{yy}}{\sigma_0} & =  -\frac{b'_yx}{R_a(R_a + \lambda)},                                                                                                                                               \\
        \frac{\bar{\sigma}_{zz}}{\sigma_0} & =  - b'_y \left\{ \frac{2\nu x}{R_a(R_a + \lambda)} \left[ 1 + \frac{a^2}{2R_a^2} + \frac{a^2}{2R_a(R_a + \lambda)} \right] + \frac{x\lambda}{R_a^3} \right\},                      \\
        \frac{\bar{\sigma}_{xy}}{\sigma_0} & = -\frac{b'_x x}{R_a(R_a + \lambda)},                                                                                                                                               \\
        \frac{\bar{\sigma}_{xz}}{\sigma_0} & =  b'_y \left[ -\frac{\nu}{R_a} + \frac{x^2}{R_a^3} + (1-\nu)\frac{a^2}{2R_a^3} \right],                                                                                            \\
        \frac{\bar{\sigma}_{yz}}{\sigma_0} & = b'_x \left[ \frac{\nu}{R_a} - (1-\nu)\frac{a^2}{2R_a^3} \right] - \frac{b'_z(1-\nu)x}{R_a(R_a + \lambda)} \left[1 + \frac{a^2}{2R_a^2} + \frac{a^2}{2R_a(R_a + \lambda)},\right], \\
    \end{aligned}
\end{equation}

When the field point is located right to the dislocation segment, i.e. $ z_1 < z_2 < z$, the following form 2 should be used
\begin{equation}
    \begin{aligned}
        \frac{\bar{\sigma}_{xx}}{\sigma_0} & = - \frac{b'_y x}{R_a(R_a - \lambda)} \left[1 - \frac{x^2 + a^2}{R_a^2} - \frac{x^2 + a^2}{R_a(R_a + \lambda)}\right],                                                               \\
        \frac{\bar{\sigma}_{yy}}{\sigma_0} & = \frac{b'_y x \lambda}{\rho_a^2 R_a},                                                                                                                                               \\
        \frac{\bar{\sigma}_{zz}}{\sigma_0} & =  b'_y \left\{ \frac{2\nu x}{R_a(R_a - \lambda)} \left[ 1 + \frac{a^2}{2R_a^2} + \frac{a^2}{2R_a(R_a - \lambda)} \right] + \frac{x\lambda}{R_a^3} \right\},                         \\
        \frac{\bar{\sigma}_{xy}}{\sigma_0} & = \frac{b_x x}{R_a(R_a - \lambda)},                                                                                                                                                  \\
        \frac{\bar{\sigma}_{xz}}{\sigma_0} & = b'_y \left[ -\frac{\nu}{R_a} + \frac{x^2}{R_a^3} + (1-\nu)\frac{a^2}{2R_a^3} \right],                                                                                              \\
        \frac{\bar{\sigma}_{yz}}{\sigma_0} & =  b'_x \left[ \frac{\nu}{R_a}  - (1-\nu)\frac{a^2}{2R_a^3} \right] + \frac{b'_z(1-\nu)x}{R_a(R_a - \lambda)} \left[1 + \frac{a^2}{2R_a^2} + \frac{a^2}{2R_a(R_a - \lambda)}\right], \\
    \end{aligned}
\end{equation}
where
\begin{align}
    \rho_a = \sqrt{x^2 + y^2 + a^2}.
\end{align}

Finally, when the field point is located between the end points of the dislocation segment, i.e. $z_1 \le z \le z_2$, the following form 3 should be used

\begin{equation}
    \begin{aligned}
        \frac{\bar{\sigma}_{xx}}{\sigma_0} & = - \frac{b'_y x \lambda}{\rho_a^2 R_a} \left[ 1 - \frac{2(x^2 + a^2)}{\rho_a^2} - \frac{x^2+a^2}{R_a^2} \right],                                                \\
        \frac{\bar{\sigma}_{yy}}{\sigma_0} & = \frac{b'_y x \lambda}{\rho_a^2 R_a},                                                                                                                           \\
        \frac{\bar{\sigma}_{zz}}{\sigma_0} & =  b'_y \left\{ \frac{2\nu x\lambda}{\rho_a^2R_a} \left[ 1 + \frac{a^2}{\rho_a^2} + \frac{a^2}{2R_a^2} \right] + \frac{x\lambda}{R_a^3} \right\},                \\
        \frac{\bar{\sigma}_{xy}}{\sigma_0} & = \frac{b_x x\lambda}{\rho_a^2R_a} \left[1 - \frac{2y^2}{\rho_a^2} - \frac{y^2}{R_a^2} \right],                                                                  \\
        \frac{\bar{\sigma}_{xz}}{\sigma_0} & =  b'_y \left[ -\frac{\nu}{R_a} + \frac{x^2}{R_a^3} + (1-\nu)\frac{a^2}{2R_a^3} \right],                                                                         \\
        \frac{\bar{\sigma}_{yz}}{\sigma_0} & = b'_x \left[ \frac{\nu}{R_a} - (1-\nu)\frac{a^2}{2R_a^3} \right] + \frac{b'_z(1-\nu)x}{\rho_a^2R_a} \left[1 + \frac{a^2}{\rho_a^2} + \frac{a^2}{2R_a^2}\right], \\
    \end{aligned}
\end{equation}
Several typos in the stress expressions in~\citep{Cai2006} have been corrected in the above.

\addcontentsline{toc}{section}{References}
\bibliographystyle{model1-num-names}

\end{document}